\documentclass[11pt,a4paper]{article}
\usepackage{jcappub}

\usepackage{amsmath,amsfonts,amssymb}
\usepackage{mathrsfs}
\usepackage{graphicx}
\usepackage[english]{babel} 
\usepackage{color}
\usepackage{bbold}
\usepackage{adjustbox}
\usepackage[utf8]{inputenc} 
\usepackage[colorlinks=true,citecolor=blue,urlcolor=blue]{hyperref}

\def\OP#1{{\color{magenta} #1}}
\def\parfrac#1#2{{\left(\frac{#1}{#2}\right)}}

\begin{document}
	\subheader{CERN-TH-2022-214}
	\title{Gravitational Waves from Domain Walls in Pulsar Timing Array Datasets} 
	
	\author[a]{Ricardo Z.~Ferreira}
        									\author[b]{Alessio~Notari}
                                       
                                            \author[a]{Oriol~Pujolàs}
                                            \author[c]{and Fabrizio~Rompineve}
                                            \affiliation[a]{Institut de F\'isica d'Altes Energies (IFAE) and The Barcelona Institute of Science and Technology (BIST), \\
		Campus UAB, 08193 Bellaterra, Barcelona, Spain
		\looseness=-1}
                                            \affiliation[b]{Departament de F\'isica Qu\`antica i Astrofis\'ica \& Institut de Ci\`encies del Cosmos (ICCUB), Universitat de Barcelona, Mart\'i i Franqu\`es 1, 08028 Barcelona, Spain
		\looseness=-1}
                                            \affiliation[c]{CERN, Theoretical Physics Department, 1211 Geneva 23, Switzerland}
                                            \emailAdd{rzambujal@ifae.es}
                                            \emailAdd{notari@fqa.ub.edu}
                                            \emailAdd{pujolas@ifae.es}
                                            \emailAdd{fabrizio.rompineve@cern.ch}

	\abstract{ We present a model-independent  search for a gravitational wave background from cosmic domain walls (DWs) in the NANOGrav 12.5 years dataset and International PTA Data Release 2. DWs that annihilate at temperatures $\sim 20-50~\text{MeV}$ with tensions $\sim (40-100~\text{TeV})^3$ provide as good a fit to both datasets as the astrophysical background from supermassive black hole mergers. DWs may decay into the Standard Model (SM) or a dark sector. In the latter case we predict an abundance $\Delta N_{\text{eff}}$ of dark radiation well within the reach of upcoming CMB surveys. Complementary signatures at colliders and laboratories can arise if couplings to the SM are present. As an example, we discuss heavy axion scenarios, where DW annihilation may interestingly be induced by QCD confinement.}

	\maketitle

\section{Introduction} 

Pulsar Timing Arrays (PTAs) are currently the most sensitive Gravitational Wave (GW) observatories and can detect a stochastic background of GWs (GWB) at frequencies of $1-100~\text{nHz}$. In the last two years, all operative PTAs (NANOGrav~\cite{Brazier:2019mmu}, European PTA~\cite{Desvignes:2016yex} and Parkes PTA~\cite{Kerr:2020qdo}) have reported strong evidence for a common-spectrum process in their searches for a GWB~\cite{NANOGrav:2020bcs, Goncharov:2021oub, Chen:2021rqp}. This has been recently reinforced by the International PTA (IPTA) analysis of older data releases from individual PTAs~\cite{Antoniadis:2022pcn}. Conclusive evidence for a GWB requires the emergence of spatial ``Hellings-Downs'' correlations~\cite{Hellings:1983fr}, which the NANOGrav (NG) collaboration expects to detect near its 18 years baseline~\cite{NANOGrav:2020spf}, if the current signal is indeed due to GWs. 

Under this assumption, a common interpretation of the signal is of astrophysical nature, as the GWB generated by Supermassive Black Hole Binaries (SMBHBs)~\cite{Burke-Spolaor:2018bvk}. Alternatively, it could originate from violent processes in the Early Universe, due to physics beyond the Standard Model (SM). In the latter case, PTAs frequencies correspond to the inverse Hubble radius at the interesting range of temperatures around the QCD phase transition (PT), i.e. $10~\text{MeV}-1~\text{GeV}$ (assuming standard cosmology since that epoch).

This far-reaching interpretation has been motivating searches in PTA datasets~\cite{Bian:2020urb, NANOGrav:2021flc, Xue:2021gyq, Wang:2022wwj,Wang:2022rjz}, mostly dedicated to first order PTs. In this paper, we focus instead on the GWB from cosmic domain walls (DWs), which we search for in the NG 12.5 years~\cite{NANOGrav:2020gpb} (NG12) and IPTA Data Release 2~\cite{Perera:2019sca} (IPTADR2) datasets (PPTA~\cite{Goncharov:2021oub} and EPTA~\cite{Chen:2021rqp} datasets lead to very good agreement with IPTADR2, see~\cite{Antoniadis:2022pcn}, thus their inclusion would not significantly alter our conclusions). DWs are topological defects that form when a discrete symmetry is spontaneously broken (see e.g.~\cite{Vilenkin:2000jqa}), leaving different Hubble patches in different degenerate vacua. When the breaking occurs after inflation, DW networks can generically be long-lived and exhibit a so-called scaling behavior~\cite{Zeldovich:1974uw, Kibble:1976sj}, independently of initial conditions and of the detailed particle physics origin. 
GWs are continuously sourced by the motion of the network until its final decay, whose occurrence is an observational requirement~\cite{Zeldovich:1974uw}.
Due to their scaling properties, DWs are very efficient sources of GWs and are thus an especially interesting particle physics target for PTAs~(in the context of NG12, see~\cite{Craig:2020bnv, Chiang:2020aui, Sakharov:2021dim} for previous work and~\cite{Ellis:2020ena, Blasi:2020mfx, Blanco-Pillado:2021ygr, Vaskonen:2020lbd, Nakai:2020oit, Neronov:2020qrl} for other scenarios). Furthermore, they arise in well-motivated particle physics frameworks, such as: two~\cite{Battye:2020jeu, Arkani-Hamed:2020yna}, twin~\cite{Chacko:2005pe, Batell:2019ptb} and composite~\cite{DiLuzio:2019wsw,Panico:2015jxa} Higgs models, non-Abelian flavor symmetries \cite{Riva:2010jm, Gelmini:2020bqg},  axion monodromy  \cite{Hebecker:2016vbl, Krajewski:2021jje}, supersymmetry \cite{Witten:1982df,Ellis:1986mq,Abel:1995wk,Dvali:1996xe,Kovner:1997ca}, grand unification \cite{Lazarides:1981fv, Everett:1982nm, Lazarides:1982tw}, discrete spacetime symmetries \cite{Dashen:1970et,Rai:1992xw,Smilga:1998dh,Gaiotto:2017tne,Craig:2020bnv}.

To encompass such a variety of possible origins, in this work we perform  model independent searches, while also obtaining specific results for $\mathbb{Z}_N$ symmetries embedded in $U(1)$ global transformations~\cite{Lazarides:1981fv, Vilenkin:1982ks}, of the type arising e.g. from axions~\cite{Vilenkin:2000jqa}. Our work presents important novelties compared to previous searches for primordial sources~\cite{Bian:2020urb, NANOGrav:2021flc, Xue:2021gyq, Wang:2022wwj, Wang:2022rjz}: (a) we perform model-independent analyses in both the NG12 and IPTADR2 datasets;  (b) we properly account for cosmological constraints and include the temperature dependence of the number of relativistic degrees of freedom in the plasma (according to~\cite{Borsanyi:2016ksw}); (c) we discuss a particle physics realization with signatures at colliders and other experiments. Additionally, compared to~\cite{Bian:2020urb, Wang:2022rjz}, we include the GWB from SMBHBs in the search for the DW signal.

\section{Gravitational waves from DWs} 

In the absence of significant interactions with the surrounding plasma, a generic DW network that forms after inflation quickly reaches a scaling regime with energy density~\cite{Vilenkin:2000jqa} 
\begin{eqnarray}
\label{eq:dwen}
	\rho_{\text{\tiny DW}}= c\,\sigma H,
\end{eqnarray}
and with $O(1)$ walls of size $H^{-1}$ moving at relativistic speed, where $H$ is the Hubble rate,  $c\sim O(\text{few})$ is a model-dependent prefactor and $\sigma$ is the DW surface energy density, or tension. In scalar field models, the DW width is of the order of the inverse mass $m$ of the constituent field.

According to \eqref{eq:dwen}, DWs tend to rapidly overclose the Universe~\cite{Zeldovich:1974uw}. This can however be avoided if the network starts annihilating at some time $t_\star$~\cite{Vilenkin:1981zs, Lazarides:1981fv, Sikivie:1982qv, Vilenkin:1982ks}. Assuming radiation domination, i.e. $H(T)=\sqrt{\pi^2 g_{*}/90}~T^2/M_p$ with $M_p\equiv (8\pi G)^{-1/2}$, the fraction of the total energy density in DWs at $t_\star$ is
\begin{equation}
\label{eq:alphann}
\alpha_{\star} \equiv  \frac{\rho_{\text{\tiny DW}}}{3 H^2 M_p^2}\Big|_\star\simeq c\, \sqrt{\frac{g_{*}(T_{\star})}{10.75}}\left(\frac{\sigma^{1/3}}{10^{5}\,\text{GeV}}\right)^3\left(\frac{10\,\text{MeV}}{T_{\star}}\right)^{2}
\end{equation}
 where ($g^s_{*}(T_{\star})$)~$g_{*}(T_{\star})$ is the number of (entropic) relativistic degrees of freedom (we approximate $g_{*,s}=g_{*}$) at the annihilation temperature $T_{\star}$.  
The temperature normalization roughly corresponds to the region preferred by the data (see below). The normalization for $\sigma^{1/3}$ then corresponds to an upper limit on the DW tension for annihilation around this temperature.

The network has a large time-varying quadrupole that efficiently radiates GWs~\cite{Vilenkin:1981zs, Preskill:1991kd, Chang:1998tb, Gleiser:1998na, Hiramatsu:2013qaa}, with a fraction $\rho_{\text{\tiny GW}}/(3H^2M_p^2)\sim 3/(32\pi)\,\alpha^2\, $ of the total energy density that is maximal at $T_{\star}$. Most GWs are emitted at a frequency $f_p\simeq H$, the inverse length of the walls in the scaling regime, that redshifted to today corresponds to: \begin{eqnarray} \label{eq:pfreq}
	f_p^0\simeq 10^{-9} \, \text{Hz} \, \parfrac{g_{*}(T_{\star})}{10.75}^{\frac{1}{6}} \parfrac{T_\star}{10\,\text{MeV}}\, .
\end{eqnarray}
The relic abundance today, $\Omega_\text{\tiny GW,DW}(f)\equiv d\rho_{\text{\tiny GW}}/d\log(f)/(3H_0^2M_p^2)$, can then be expressed as
\begin{align}
\label{eq:omegadw}
&\Omega_\text{\tiny GW,DW}(f)h^2\simeq 10^{-10}\, \tilde{\epsilon}\left(\frac{10.75}{g_{*}(T_{\star})}\right)^{\frac{1}{3}}\left(\frac{\alpha_{\star}}{0.01}\right)^{2} S\left(\frac{f}{f_{p}^{0}}\right),
\end{align}
where $\tilde{\epsilon}\simeq 0.1-1$ is an efficiency parameter to be extracted from numerical simulations~\cite{Hiramatsu:2013qaa}
and the function $S(x)$ describes the spectral shape of the signal. A useful parametrization (see e.g.~\cite{Caprini:2019egz}) is:
\begin{equation}
\label{eq:spectrum}
S(x)= \frac{(\gamma+\beta)^\delta}{(\beta x^{-\frac{\gamma}{\delta}} + \gamma x^{\frac{\beta}{\delta}})^\delta},
\end{equation}
where $\gamma, \beta$ capture the spectral slopes at $f\ll f_{p}^0$ and $f\gg f_{p}^0$ respectively, and $\delta$ the width around the maximum. While $\gamma=3$ because of causality (e.g.~\cite{Caprini:2019egz}),  numerical analyses are needed to determine $\delta$ and $\beta$. Most recent simulations~\cite{Hiramatsu:2013qaa} find $\delta, \beta \simeq 1$, although results for $\mathbb{Z}_N$ 
hybrid string-wall networks suggest that $\beta$ decreases with increasing $N$~\cite{Hiramatsu:2012sc}. The spectrum is cut off at frequencies larger than the (redshifted) inverse of the wall width $\sim m$. We stress that the above estimates only account for emission during the scaling regime. The subsequent annihilation of the network may further source GWs if sufficiently violent, but we neglect such contributions here since they are model-dependent and have not yet been numerically investigated.~\footnote{
The GW signal is suppressed when DWs interact with the plasma down to $T_{\star}$~\cite{Vilenkin:2000jqa, Nakayama:2016gxi}, never achieving the scaling regime. The spectral shape may differ from~\eqref{eq:spectrum} if the DW network decays because of symmetry restoration~\cite{Vilenkin:1981zs, Babichev:2021uvl}.}

Our discussion has so far been independent of the specific DW annihilation mechanism, and so will be most of the results presented in this work. Additionally, we will consider a well-studied annihilation mechanism in more detail: explicit symmetry breaking by a tiny (vacuum) energy density gap $\Delta V$ between vacua~\cite{Sikivie:1982qv, Gelmini:1988sf} (see e.g.~\cite{Avelino:2008mh} for other possibilities). This results in a pressure $p\sim \Delta V$. The annihilation temperature can then be estimated using $\Delta V\simeq \rho_{\text{\tiny DW}}$:
\begin{equation}
\label{eq:Tann}
T_{\star}\simeq \frac{5\,\text{MeV}}{\sqrt{c}}\left(\frac{10.75}{g_{*}(T_{\star})}\right)^{\frac{1}{4}}\left(\frac{\Delta V^{1/4}}{10\,\text{MeV}}\right)^2
\left(\frac{10^5\,\text{GeV}}{\sigma^{1/3}}\right)^{\frac32}.
\end{equation}
For consistency with the GW estimates above, we neglect the contribution of $\Delta V$ to the energy density in the DW network, which is at most comparable to $\rho_{\text{\tiny DW}}$. We thus see that the typical scale of the energy gap suggested by the data is $\gtrsim 10~\text{MeV}$.

Overall, the GW signal from DWs depends only on $T_{\star}$ and $\alpha_{\star}$. Given the current uncertainties in the determination of $\delta, \beta$ from simulations, we also consider slight deviations from their reference unit value, thereby allowing for a total of four parameters in our analysis of GWs from DWs.  In models with a gap, it is useful to replace $T_{\star}$ and $\alpha_{\star}$ with the DW tension $\sigma$ and $\Delta V$, by means of~\eqref{eq:alphann},~\eqref{eq:Tann}. 

\section{Cosmology} Most energy density from DW annihilation is typically released in mildly (or non-) relativistic quanta of the wall constituents~\cite{Kawasaki:2014sqa}.  When these are stable, they dilute as matter and rapidly dominate over the radiation background, since the DWs of interest make up a significant fraction of the total energy density at $T_\star$, thereby spoiling Big Bang Nucleosynthesis (BBN) and Cosmic Microwave Background (CMB) anisotropies. Thus, they must decay to dark radiation or \OP{to} SM particles. We consider such decays to be efficient at $T_{\star}$, which leads to the weakest constraints on the DW interpretation of PTA data.

\begin{figure}
    \begin{center}
        \includegraphics[width=0.45\textwidth]{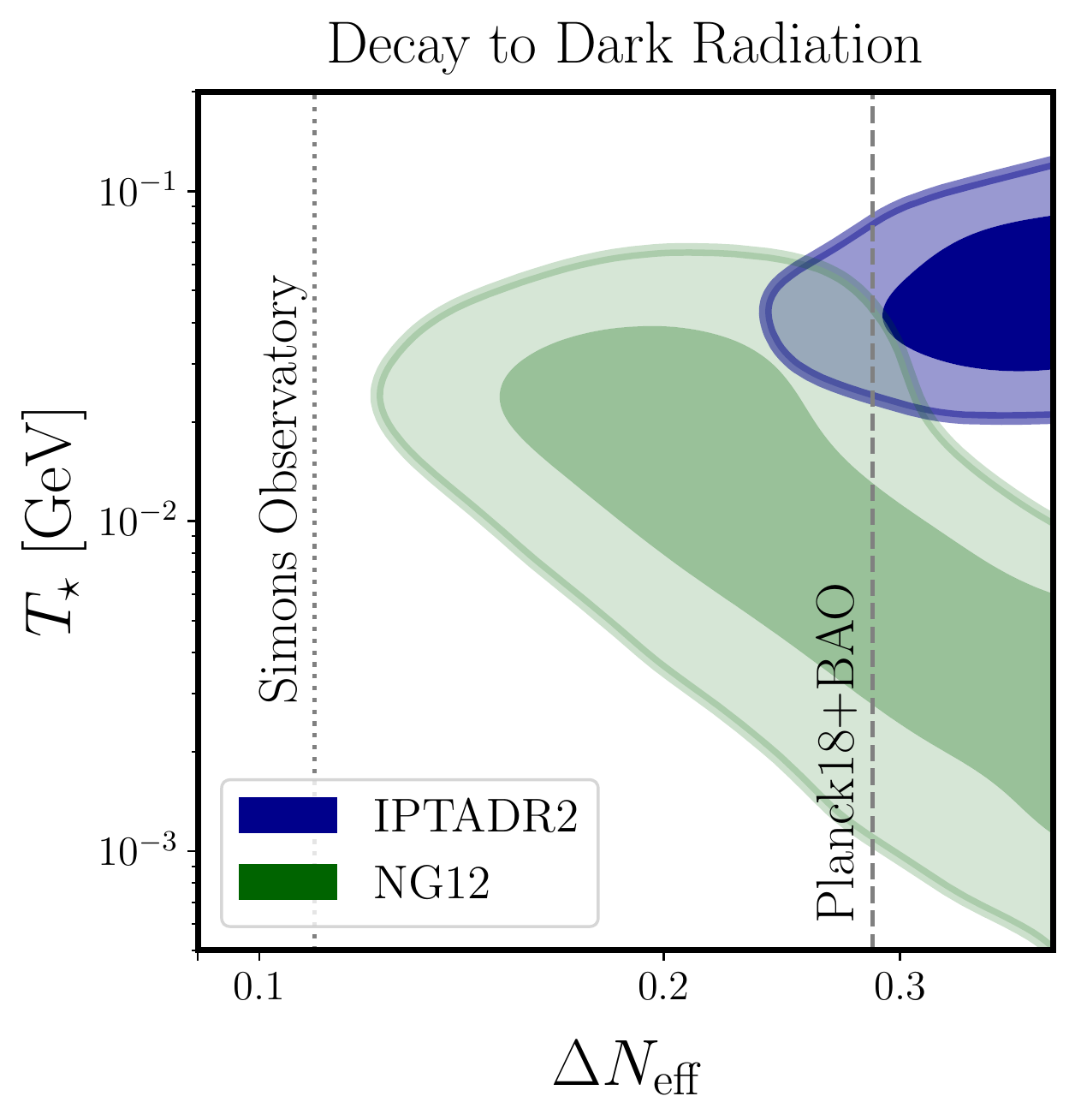}
        \hfill
        \includegraphics[width=0.45\textwidth]{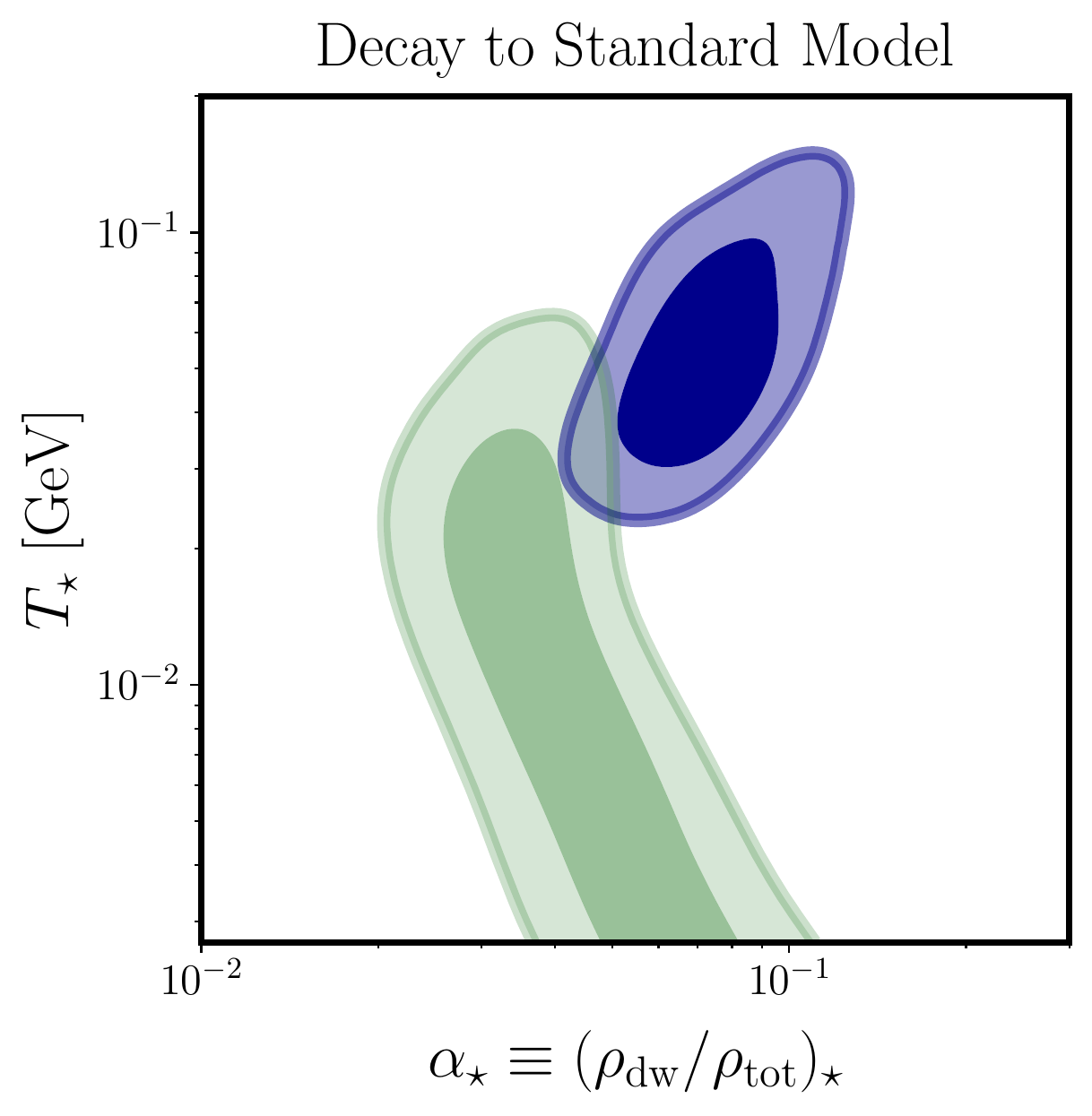}
        
    \end{center}
        	\caption{\small $1\sigma$ and $2\sigma$ contours for the 2d posterior distributions of DW parameters. {\it Left panel:} DW constituents decay to dark radiation. In this case, the prior $\Delta N_{\text{eff}}\leq 0.39$ from BBN+$Y_p$+D ($95\%$ C.L.~\cite{Fields:2019pfx}) is applied, as well as $T_\star\geq 500~\text{keV}$. The $95\%$ C.L. bound from Planck18+BAO~\cite{Planck:2018vyg} and $95\%$ C.L. forecasted reach of Simons Observatory~\cite{SimonsObservatory:2018koc} are shown as dashed and dotted lines respectively. {\it Right panel:} DW constituents decay to SM radiation. The priors $\alpha_\star\leq 0.3$ and $T_\star\geq 2.7~\text{MeV}$ are applied. See Appendix~\ref{sec:priors} for 1d and 2d posteriors of all parameters.}
        \label{fig:NGDW}
\end{figure}

We first discuss the scenario of decay into dark radiation (henceforth, the DR scenario). The abundance of DR is commonly expressed as the effective number of neutrino species $\Delta N_{\text{eff}}\equiv \rho_\text{\tiny DR}/\rho_{\nu}$ ($\rho_\nu$ being the energy density of a single neutrino species). The constraint from BBN (CMB+BAO) reads $\Delta N_{\text{eff}}\leq 0.39~\text{\cite{Fields:2019pfx}}~(0.29~\text{\cite{Planck:2018vyg}})$ at $95\%$ C.L. Setting $\rho_\text{\tiny DR}\simeq \rho_{\text{\tiny DW}}$ at $T_\star$, gives 
\begin{equation}
    \Delta N_{\text{eff}}\simeq 13.6 \, g_{*}(T_{\star})^{-1/3}\alpha_\star,
\end{equation} 
thus the bounds above translate to $\alpha_\star \lesssim 0.06~(0.05)\times 10.75/g_{*}(T_{\star})$. We do not consider the case of DW decay after $e^{+}e^{-}$ annihilation (i.e. $T_{\star}\lesssim 500~\text{keV}$), nor the case of DW constituents diluting as matter and decaying after $T_{\star}$, as both cases would lead to a larger $\Delta N_{\text{eff}}$.

On the other hand, when decay occurs to SM particles (henceforth, the SM scenario), BBN imposes $T_{\star}\gtrsim 2.7~\text{MeV}$ for any relevant value of $\alpha_{\star}$~\cite{Jedamzik:2006xz, Bai:2021ibt}. We also cautiously impose $\alpha_\star\leq 0.3$ to avoid deviations from radiation domination, which require dedicated numerical studies. This also ensures that the GWs emitted from DWs respect the aforementioned DR bound, since $\Delta N_{\text{eff, gw}}\simeq 0.2\, \alpha_\star^2 (g_{*}(T_{\star})/10.75)^{-1/3}$, see~\eqref{eq:omegadw}.

\section{Data Analysis} GW searches at PTAs are performed in terms of the timing-residual cross-power spectral density $S_{ab}(f)\equiv \Gamma_{ab} h^2_c(f)/(12\pi^2)f^{-3}$, where $h_c(f)\simeq 1.26\cdot 10^{-18}(\text{Hz}/f)\sqrt{h^2\Omega_{\text{GW}}}$ (see e.g.~\cite{Caprini:2018mtu}) is the characteristic strain spectrum and $\Gamma_{ab}$ contains correlation coefficients between pulsars $a$ and $b$ in a given PTA. We performed Bayesian analyses using the codes {\tt enterprise}~\cite{enterprise} and {\tt enterprise\_extensions}~\cite{enterprise_ext}, in which we implemented the DW signal \eqref{eq:omegadw},\eqref{eq:pfreq},\eqref{eq:spectrum}, and {\tt PTMCMC}~\cite{justin_ellis_2017_1037579} to obtain MonteCarlo samples. We derive posterior distributions using {\tt GetDist}~\cite{Lewis:2019xzd}. We include white, red and dispersion measures noise parameters following the choices of the NG12~\cite{NANOGrav:2020bcs} and IPTADR2~\cite{Antoniadis:2022pcn} searches for a common spectrum. Furthermore, we limit the stochastic GW search to the lowest 5 and 13 frequencies of the NG12 and IPTADR2 datasets respectively to avoid pulsar-intrinsic excess noise at high frequencies, as in~\cite{NANOGrav:2020bcs, Antoniadis:2022pcn}. We fix $\tilde{\epsilon}=0.7$ according to~\cite{Hiramatsu:2013qaa} and discuss different choices below. Further details and prior choices are reported in Appendix~\ref{sec:priors}.

We first obtain results with DWs as the only source of GWs and separately analyze the DR and SM scenarios. In the former case, we sample $\Delta N_{\text{eff}}$ and $T_{\star}$ logarithmically, $\Delta N_{\text{eff}}\in [10^{-2}, 0.39]$,  $T_{\star}\in [5\cdot 10^{-4}, 10]~\text{GeV}$. For the SM scenario, we trade $\Delta N_{\text{eff}}$ for $\alpha_{\star}\in [10^{-3}, 0.3]$ and impose $T_{\star}\geq 2.7~\text{MeV}$. In all analyses we sample $\beta\in [0.5, 1]$ and $\delta\in [0.3, 3]$. 

Posterior distributions are shown in Fig.~\ref{fig:NGDW}. In both scenarios, NG12 is well fitted by the high frequency tail of the spectrum, i.e. by a simple power law ($\beta=1$ or $\gamma=6$ in the notation of~\cite{NANOGrav:2020bcs}). On the other hand, IPTADR2~\cite{Antoniadis:2022pcn} prefers the region of the spectrum around the peak. We find almost flat posteriors for $\beta$ and $\delta$, see Appendix~\ref{sec:priors}.

\begin{figure}[t]
\centering
\includegraphics[width=0.44\textwidth]{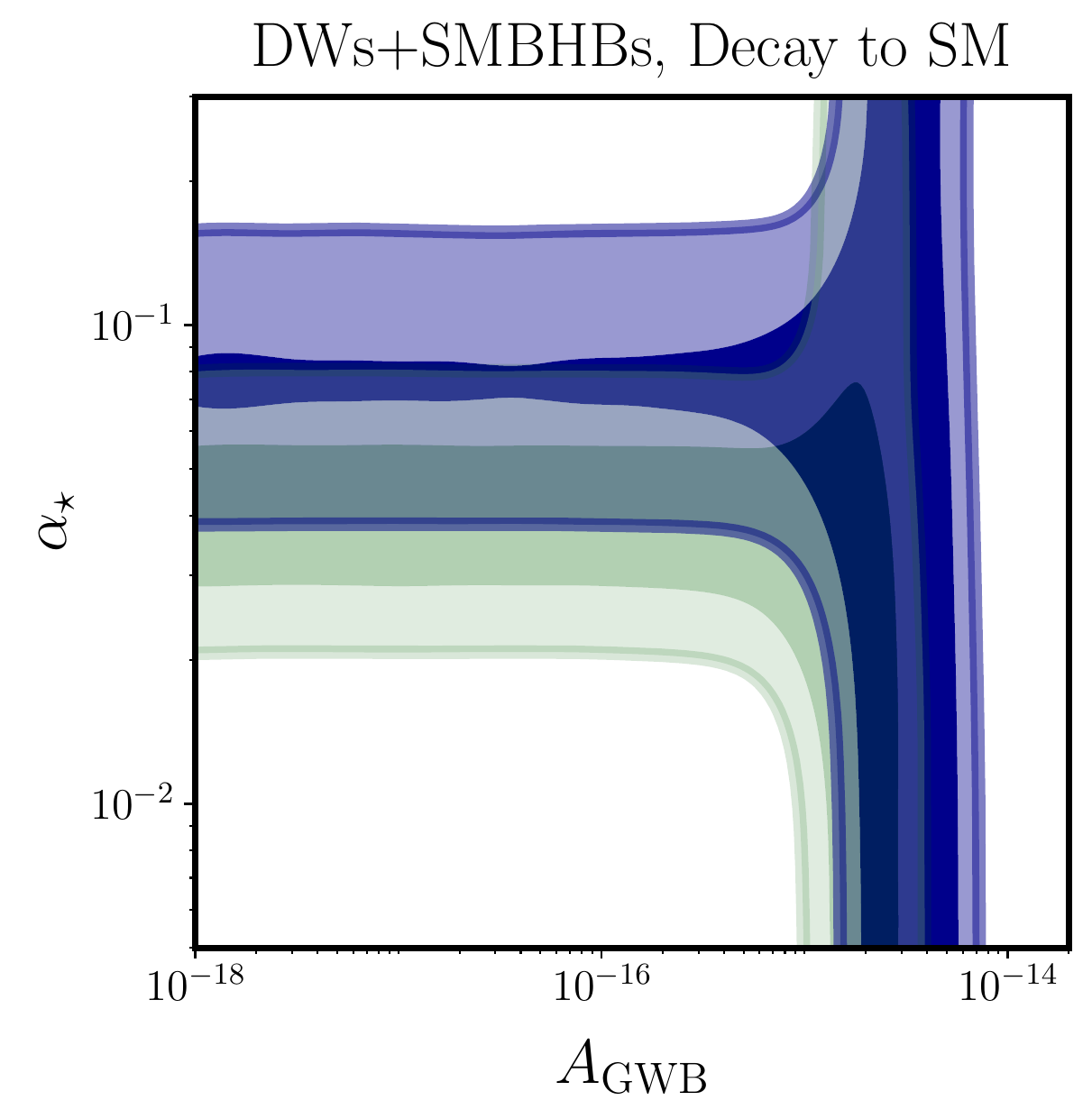}
\hfill
\includegraphics[width=0.45\columnwidth]{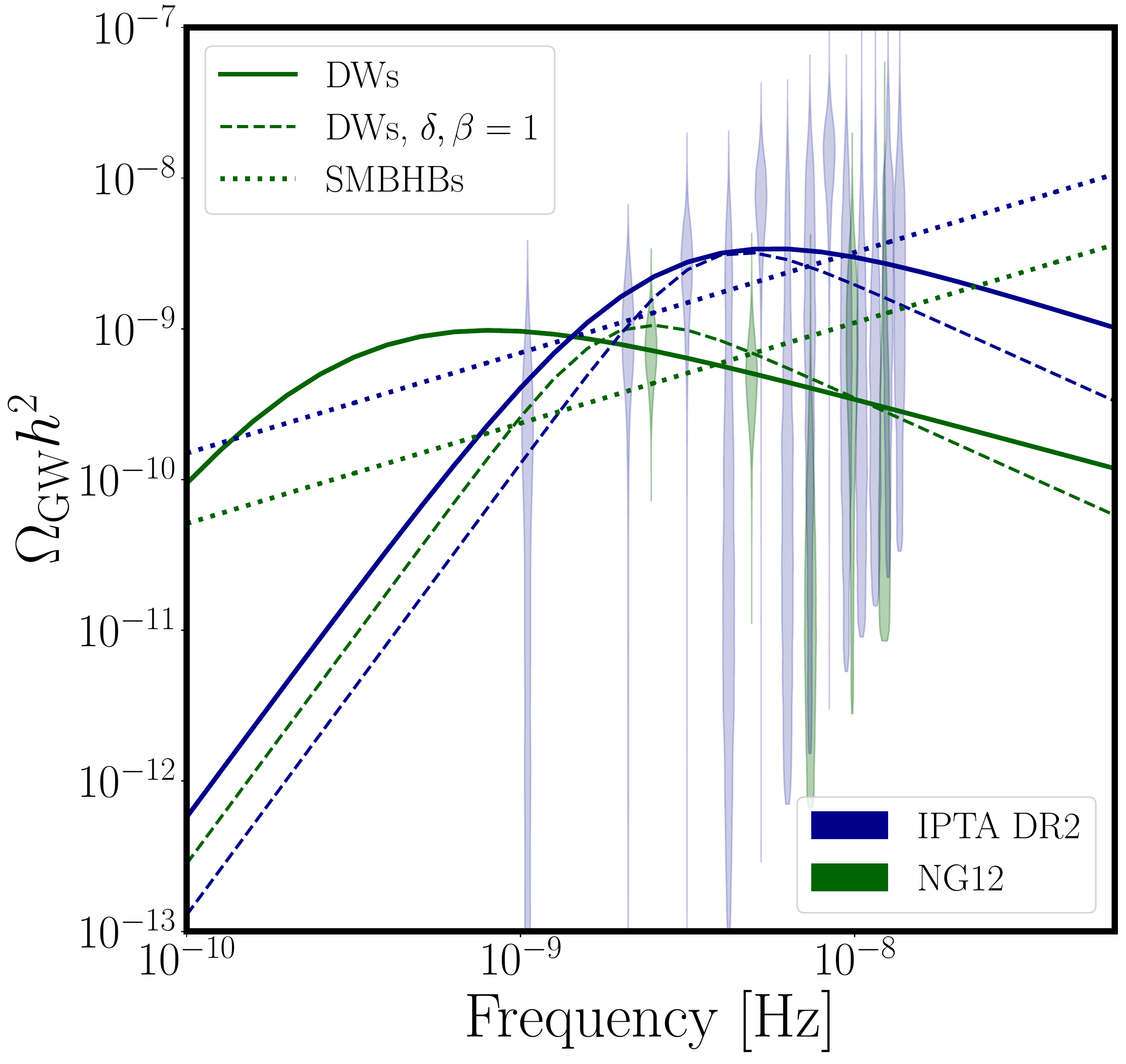}
\caption{\small {\it Left  panel:} including a GW background (GWB) from SMBHBs, with amplitude $A_{\text{GWB}}$, fixing $\tilde{\epsilon}=0.7$ (see text for different choices).  {\it Right panel:}   Maximum likelihood GW abundance from DWs (solid, dashed), assuming the SM scenario, and from SMBHBs (dotted). For comparison, we show the free spectrum posteriors obtained by converting the results of~\cite{NANOGrav:2020bcs} (NG12) and~\cite{Antoniadis:2022pcn} (IPTA DR2) (violin shapes, lower limits set by priors). Solid lines correspond to $\alpha_\star\simeq 0.07\, (0.04), T_\star\simeq 47\, (7)~\text{MeV}, \delta\simeq 2 \,(2.5), \beta\simeq 0.7\, (0.6)$ for IPTA DR2 (NG12), dashed lines to $\delta, \beta = 1, \alpha_\star\simeq 0.07\, (0.04), T_\star\simeq 42\, (21)~\text{MeV}$.  $A_{\text{GWB}}\simeq 3.3\,(1.8)\cdot 10^{-15}$, according to our IPTA DR2 (NG12) DWs+SMBHBs analysis. See Appendix~\ref{sec:priors} for 1d and 2d posteriors of all parameters.}
\label{fig:bestfit}
\end{figure}

For the DR scenario, Fig.~\ref{fig:NGDW} (left), a significant portion of the parameter space is constrained by the BBN prior. We find $\Delta N_{\text{eff}}\geq 0.26~(0.15)$ at $95\%$ C.L. from IPTADR2 (NG12). These values are close to the current bound from Planck18+BAO (dashed line, $2\sigma$) and well within the reach of the upcoming Simons Observatory~\cite{SimonsObservatory:2018koc} (dotted line, $2\sigma$). However, note that CMB bounds only apply if the decay products remain relativistic until recombination. We also find $T_\star\in [23, 93]~(\leq 51)~\text{MeV}$ at $95\%$ C.L. from IPTADR2 (NG12).

For the SM scenario, Fig.~\ref{fig:NGDW} (right), we find $\alpha_\star \in [0.05, 0.11]~([0.02, 0.08])$, well below the upper prior boundary, and $T_\star \in [27, 121]~(\leq 41)~\text{MeV}$ at $95\%$ C.L. from IPTADR2 (NG12). Further details and posteriors can be found in Appendix~\ref{sec:priors}.

Next, we search for GWs from DWs in the presence of a stochastic background from SMBHBs, whose strain we take to be given by the simple power law $h_c(f)=A_{\text{GWB}}(f/\text{yr}^{-1})^{-2/3}$ (see e.g.~\cite{Burke-Spolaor:2018bvk}), assuming the SM scenario.\footnote{The spectral tilt of the GWB from SMBHBs can differ from the standard value considered above (and in~\cite{NANOGrav:2020bcs, Antoniadis:2022pcn}), see e.g.~\cite{Middleton:2020asl} for a recent analysis in light of NG12. We notice that the choice above is in excellent agreement with the IPTADR2 results on the tilt of the power-law spectrum, whereas a flatter spectrum would be slightly preferred by NG12. Therefore, we expect our results with the IPTADR2 dataset not to be significantly affected by the inclusion of the spectral tilt as an extra parameter, whereas the NG12 results may be slightly altered (the GWB from SMBHBs may become even more degenerate with the low-frequency tail of the DW signal). In any case, we will show below that the NG12 dataset does not show preference for any of the two models against the other one, even with the choice above.} The 2D posterior distribution of $\alpha_\star$ and $A_{\text{GWB}}$ are reported in Fig.~\ref{fig:bestfit} (left panel). In particular, the central values of $A_{\text{GWB}}$ agree with~\cite{NANOGrav:2020bcs, Antoniadis:2022pcn}, and we find broader posteriors due to our additional background from DWs. When the PTA excess is mostly modeled by SMBHBs, the DW parameter $\alpha_{\star}$ is only limited by our priors and can be large when the peak of the spectrum from DWs is located out of the PTA sensitivity band. We compare models using the Bayes factors $\log_{10} B_{i,j}$ of model $j$ with respect to model $i$. For NG12, we find: $\log_{10} B_{\text{\tiny SMBHBs, DW}}\simeq 0.16$, $\log_{10} B_{\text{\tiny DW, DW+SMBHBs}}\simeq 0.07$. For IPTADR2, we find: $\log_{10} B_{\text{\tiny DW, SMBHBs}}\simeq 0.48$, $\log_{10} B_{\text{\tiny DW, DW+SMBHBs}}\simeq 0.38$. Thus we find no substantial evidence for one model against any other one in the datasets.

The maximum likelihood GW spectra from DWs (SM scenario), and for comparison from SMBHBs (as obtained in our DW+SMBHBs analysis), are shown in Fig.~\ref{fig:bestfit} (right panel). Spectra with $\delta, \beta=1$ are also displayed, to show the minor effect of these parameters on the quality of the fit.

\begin{figure}[t]
\centering
\includegraphics[width=0.6\columnwidth]{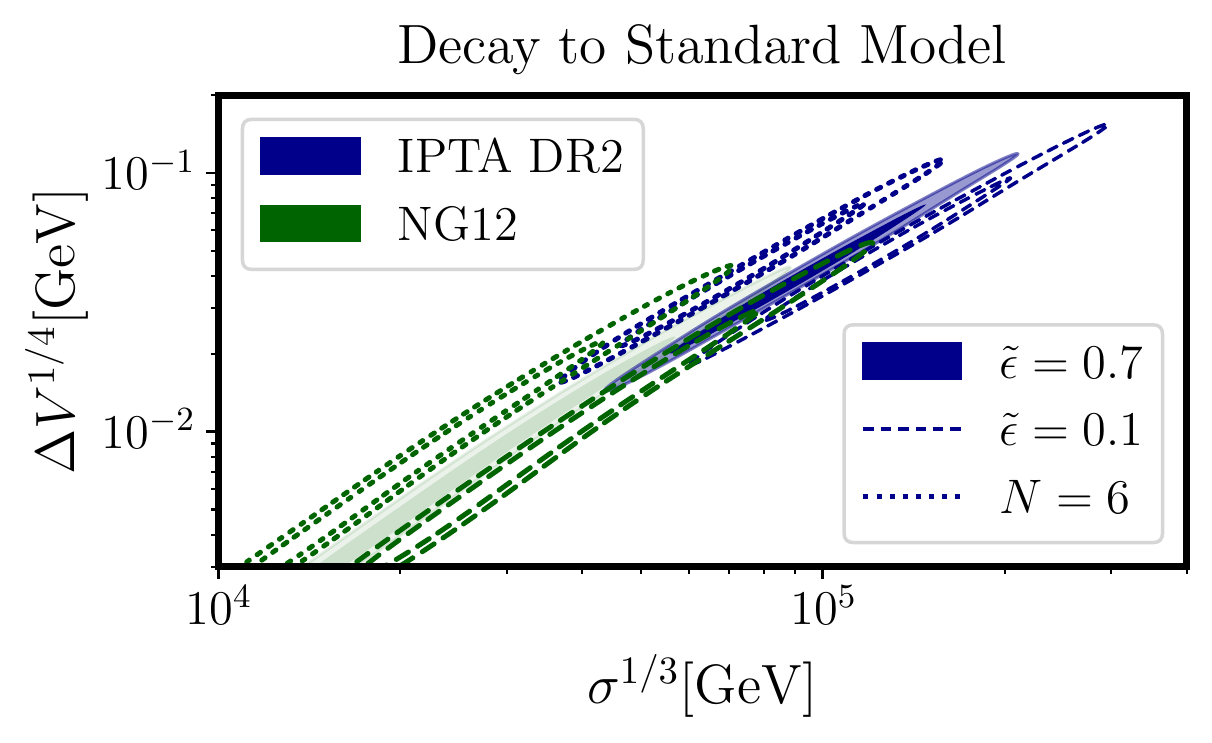}
\caption{Posterior distributions of the DW tension $\sigma^{1/3}$ and the gap energy (density) $\Delta V^{1/4}$ scales, $1\sigma$ and $2\sigma$ contours. The prior $T_{\star}\geq 2.7~\text{MeV}$ corresponds to $\Delta V^{1/4}\gtrsim 3~\text{MeV}$. See Appendix~\ref{sec:priors} for 1d and 2d posteriors of all parameters.}
\label{fig:NGDWsigma}
\end{figure}

We then specify our analysis of the SM scenario to the case of network annihilation due to a gap $\Delta V$, by sampling the tension $\sigma\in [10^{10}, 10^{18}]~\text{GeV}^3$ rather than $\alpha_{\star}$ and deriving posteriors of $\Delta V^{1/4}$ using~\eqref{eq:Tann}. We restrict our analysis to GWs from DWs only (our results will thus show the values of DW parameters which can provide a good interpretation of the data, in alternative to the SMBHBs GWB). We take $c=2.2$ (obtained from string-wall networks with $N=3$~\cite{Kawasaki:2014sqa}). Fig.~\ref{fig:NGDWsigma} shows that both datasets are well modeled when $\sigma\simeq (40-100~\text{TeV})^3$ and $\Delta V\simeq (15-50~\text{MeV})^4$.

Finally, we discuss different choices of numerical coefficients $\tilde{\epsilon}$ and $c$. From \eqref{eq:omegadw}, a change in $\tilde{\epsilon}$ can be reabsorbed by rescaling $\alpha_\star$, and thus $\Delta N_{\text{eff}}$, in Fig.~\ref{fig:NGDW}. The DR bound then severely constrains DWs as the dominant source of GWs in IPTADR2 (NG12) data, unless $\tilde{\epsilon}\gtrsim 0.3\, (0.07)$. On the other hand, imposing $\alpha_{\star}\leq 0.3$ leads to $\tilde{\epsilon}\gtrsim 0.02 (0.005)$ in the SM scenario. The effect of smaller values of $\tilde{\epsilon}$, and of larger values of $c$ is shown in Fig.~\ref{fig:NGDWsigma} (we take the example $c\simeq 4.5$ from~\cite{Kawasaki:2014sqa} for string-wall networks with $N=6$).

\section{Particle Physics Interpretations and Discussion}
Now that we have identified the properties of the DW networks that provide a good modelling of the data,
let us briefly discuss interesting microphysical origins and other potential observable signatures of such DWs.
We focus here on scenarios with DW annihilation induced by a gap $\Delta V$. Intriguingly, the preferred values of  $\Delta V$ and of the DW tension $\sigma$ shown in Fig.~\ref{fig:NGDWsigma} fall in the ballpark of two particularly interesting energy scales.

First, the $10-100~\text{TeV}$ range for $\sigma^{1/3}$ points at new physics which may be probed by (future) colliders (see e.g.~\cite{Craig:2020bnv}). Second, the $10-100~\text{MeV}$ range for $\Delta V^{1/4}$ is close to $\Lambda_{\text{QCD}}\simeq 300~\text{MeV}$, so that one may entertain the possibility of QCD inducing DW annihilation~\cite{Preskill:1991kd, Lazarides:1992gg, Higaki:2016jjh, Long:2018nsl, Chiang:2020aui}.

 A realization of the latter idea may consist of a heavy axion field $a$ with $\mathbb{Z}_N$ symmetry, $N>1$, and decay constant $F_a$, coupled to the topological term of a confining dark gauge sector $\mathcal{H}$. Upon $\mathcal{H}$-confinement at some scale $\Lambda_\mathcal{H}\gg \Lambda_\text{QCD}$, the $\mathbb{Z}_N$ symmetry is spontaneously broken and a hybrid string-wall network forms with DW tension $\sigma\simeq 8 m_a F_a^2$, where $m_a\simeq \Lambda_\mathcal{H}^2/F_a$ is the axion mass. If $a$ also couples to QCD, it receives an additional potential around the QCD PT with size set by the topological susceptibility $\Delta V_\text{QCD}^{1/4}\simeq 75~\text{MeV}$~\cite{GrillidiCortona:2015jxo}. This can induce annihilation when its periodicity differs from that of the $\mathcal{H}$-induced potential.  While a detailed exploration of this scenario is beyond the
aim of this work (see however Appendix~\ref{sec:axion}), note that solving the strong CP problem 
requires either a fine alignment between the potentials induced by $\mathcal{H}$ and QCD,
which 
might be challenging (see e.g. \cite{Dimopoulos:2016lvn,  Agrawal:2017ksf, Hook:2019qoh, Gherghetta:2020keg, ZambujalFerreira:2021cte} for recent work), or a second axion that couples only (mostly) to QCD~(see e.g.~\cite{Higaki:2016jjh, Draper:2018tmh, Long:2018nsl}). Alternatively, annihilation may occur due to (and/or in) a dark sector, see Appendix~\ref{sec:axion}.

\begin{figure}[t]
\centering
\includegraphics[width=0.5\columnwidth]{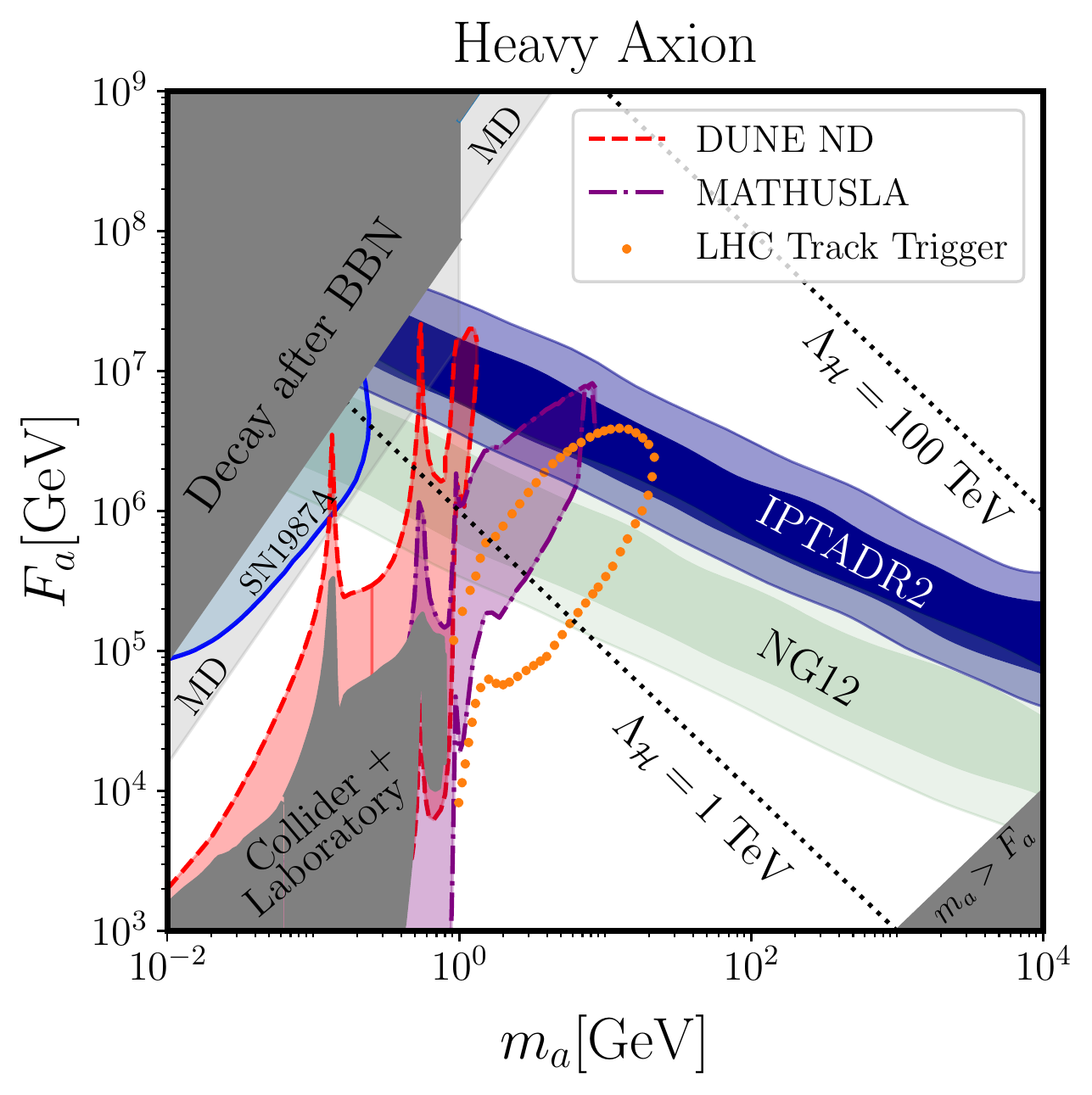}
\caption{Posterior distributions of heavy axion parameters $m_a$ and $F_a$, $1\sigma$ and $2\sigma$ contours, obtained by marginalizing over $T_{\star}\geq 2.7~\text{MeV}$. Constraints and detection forecasts are taken from~\cite{Kelly:2020dda}. In the upper left corner, the axion decays after BBN, to photons (gluons) for $m_a\lesssim (\gtrsim) \text{GeV}$~\cite{Aloni:2018vki}. Close to this region, axions radiated by DWs lead to Matter Domination (MD) before decaying altering GW predictions. The axion effective theory is not valid for $m_a>F_a$. See Appendix~\ref{sec:axion} for all posteriors.}
\label{fig:axion}
\end{figure}

We present the region of the $m_a-F_a$ parameter space for
which a heavy axion can model PTA data in
Fig.~\ref{fig:axion}, assuming decay to SM particles. Degeneracy between parameters can be
clearly observed, as the GW signal only depends on $\sigma$ and
$T_\star$. We also show existing constraints and future detection prospects from collider, astrophysics and laboratory
experiments. Remarkably, for $m_a\sim 100~\text{MeV}-20~\text{GeV}$
and $F_a\sim 10^5-2\cdot 10^7~\text{GeV}$, a heavy axion can be discovered at DUNE ND~\cite{DUNE:2020ypp} and/or MATHUSLA~\cite{Chou:2016lxi} and/or 
HL-LHC~\cite{Hook:2019qoh}, while also fitting current PTA data.

Additional observable signatures and constraints may arise from the dark confining sector, with a scale $\Lambda_\mathcal{H}\sim 1-50~\text{TeV}$ from Fig.~\ref{fig:axion} (e.g. GWs in the LISA range~\cite{Baker:2019nia} if $\mathcal{H}$ undergoes a first order PT~\cite{Schwaller:2015tja}, the presence of a dark matter candidate~\cite{Mitridate:2017oky}, or signatures at colliders~\cite{Alimena:2019zri}).

We also note that collapsing structures during DW annihilation might form primordial black holes (PBHs)~\cite{Deng:2016vzb, Ferrer:2018uiu}, whose masses depend substantially on the annihilation temperature, giving $M_\text{PBH}\sim \Delta V H_\star^{-3} \sim {\cal O}(10-10^4)~M_{\odot} $. Intriguingly, this encompasses the LIGO BH mass range. A dedicated numerical study is however required to assess the PBH abundance.

Finally, PTAs are expected to settle whether the currently observed common-process spectrum is due to GWs in the near future.
Shall this be the case, obtaining the detailed spectral shape of the GW signal from DWs, including the annihilation phase, will be crucial to distinguish it from other candidate sources. Alternatively, our work can be used to constrain scenarios with spontaneously broken discrete symmetries.

\section*{Acknowledgments}
	
	We thank Joachim Kopp for help with CERN LXPLUS cluster, on which the Bayesian analyses presented in this work were performed, and Marianna Annunziatella for suggestions on {\tt matplotlib}~\cite{4160265}, which was used for plots. The {\tt enterprise} code used in this work makes use of {\tt libstempo}~\cite{libstempo} and {\tt PINT}~\cite{Luo:2020ksx, pint}. The chain files to produce the free power spectrum violins in Fig.~\ref{fig:bestfit} were downloaded from the publicly available \href{https://zenodo.org/record/5787557}{IPTA DR2} and \href{https://data.nanograv.org/}{NG12} data releases.

This work is partly supported by the grants PID2019-108122GBC32, “Unit of Excellence Mar\'ia de Maeztu 2020-2023” of ICCUB (CEX2019-000918-M), AGAUR2017-SGR-754,  PID2020-115845GB-I00/AEI/ 10.13039/501100011033 and 2017-SGR-1069. IFAE is partially funded by the CERCA program of the Generalitat de Catalunya. 
R.Z.F. is supported by the Direcci\'o General de Recerca del Departament d’Empresa i Coneixement (DGR) and by the EC through the program Marie Sk\l odowska-Curie COFUND (GA 801370)-Beatriu de Pinos.
A.N. is grateful to the IFPU (SISSA, Trieste) for the hospitality during the course of this work.

\appendix

\section{Numerical Strategy}
\label{sec:priors}

\begin{table}
\resizebox{\textwidth}{!}{%
\begin{tabular} {| l | c| c| c| c|c|c|c|c|}
\hline\hline
 \multicolumn{1}{|c|}{ Parameter} & \multicolumn{3}{|c|}{~~~~~Description~~~~~} & \multicolumn{2}{|c|}{~~~~Prior~~~~} & \multicolumn{3}{|c|}{~~Comments~~} \\
\hline\hline
\multicolumn{9}{|c|}{\textbf{White Noise}}\\
\hline\hline
$E_k$& \multicolumn{3}{|c|}{EFAC per backend/receiver system} & \multicolumn{2}{|c|}{Uniform $[0,10]$} & \multicolumn{3}{|c|}{single-pulsar only} \\
$Q_k [s]$& \multicolumn{3}{|c|}{EQUAD per backend/receiver system} & \multicolumn{2}{|c|}{log-Uniform $[-8.5,-5]$} & \multicolumn{3}{|c|}{single-pulsar only} \\
$J_k [s]$& \multicolumn{3}{|c|}{ECORR per backend/receiver system} & \multicolumn{2}{|c|}{log-Uniform $[-8.5,-5]$} & \multicolumn{3}{|c|}{single-pulsar only (NG12, NG9)} \\
\hline\hline
\multicolumn{9}{|c|}{\textbf{Red Noise}}\\
\hline\hline
$A_{\text{red}}$& \multicolumn{3}{|c|}{Red noise power-law amplitude} & \multicolumn{2}{|c|}{log-Uniform $[-20,-11]$} & \multicolumn{3}{|c|}{one parameter per pulsar} \\
$\gamma_{\text{red}}$& \multicolumn{3}{|c|}{Red noise power-law spectral index} & \multicolumn{2}{|c|}{Uniform $[0,7]$} & \multicolumn{3}{|c|}{one parameter per pulsar} \\
\hline\hline
\multicolumn{9}{|c|}{\textbf{DM Variations Gaussian Process Noise}}\\
\hline\hline
$A_{\text{DM}}$& \multicolumn{3}{|c|}{DM noise power-law amplitude} & \multicolumn{2}{|c|}{log-Uniform $[-20,-11]$} & \multicolumn{3}{|c|}{one parameter per pulsar (IPTADR2)} \\
$\gamma_{\text{DM}}$& \multicolumn{3}{|c|}{DM noise power-law spectral index} & \multicolumn{2}{|c|}{Uniform $[0,7]$} & \multicolumn{3}{|c|}{one parameter per pulsar (IPTADR2)} \\
\hline\hline
\multicolumn{9}{|c|}{\textbf{Domain Wall Annihilation (DR scenario)}}\\
\hline\hline
$T_\star [\text{GeV}]$& \multicolumn{3}{|c|}{Annihilation temperature} & \multicolumn{2}{|c|}{log-Uniform $[\log_{10}(0.0005), 10]$} & \multicolumn{3}{|c|}{one parameter for PTA} \\
$\Delta N_{\text{eff}}$& \multicolumn{3}{|c|}{Effective number of neutrino species} & \multicolumn{2}{|c|}{log-Uniform $[-3, \log_{10}(0.39)]$} & \multicolumn{3}{|c|}{one parameter for PTA} \\
$\beta $& \multicolumn{3}{|c|}{High frequency spectral index} & \multicolumn{2}{|c|}{Uniform $[0.5, 1]$} & \multicolumn{3}{|c|}{one parameter for PTA} \\
$\delta $& \multicolumn{3}{|c|}{Spectrum width} & \multicolumn{2}{|c|}{log-Uniform $[0.3, 3]$} & \multicolumn{3}{|c|}{one parameter for PTA} \\
\hline\hline
\multicolumn{9}{|c|}{\textbf{Domain Wall Annihilation (SM scenario)}}\\
\hline\hline
$T_\star [\text{GeV}]$& \multicolumn{3}{|c|}{Annihilation temperature} & \multicolumn{2}{|c|}{log-Uniform $[\log_{10}(0.0027), 10]$} & \multicolumn{3}{|c|}{one parameter for PTA} \\
$\alpha_\star$& \multicolumn{3}{|c|}{Energy fraction in DWs} & \multicolumn{2}{|c|}{log-Uniform $[-3, \log_{10}(0.3)]$} & \multicolumn{3}{|c|}{one parameter for PTA} \\
$\beta $& \multicolumn{3}{|c|}{High frequency spectral index} & \multicolumn{2}{|c|}{Uniform $[0.5, 1]$} & \multicolumn{3}{|c|}{one parameter for PTA} \\
$\delta $& \multicolumn{3}{|c|}{Spectrum width} & \multicolumn{2}{|c|}{log-Uniform $[0.3, 3]$} & \multicolumn{3}{|c|}{one parameter for PTA} \\
\hline
$\sigma [\text{GeV}^3]$& \multicolumn{3}{|c|}{Domain Wall tension} & \multicolumn{2}{|c|}{log-Uniform $[10, 18]$} & \multicolumn{3}{|c|}{one parameter for PTA (instead of $\alpha_{\star}$)} \\
$\Delta V^{1/4}~[\text{GeV}] $& \multicolumn{3}{|c|}{Energy difference between vacua} & \multicolumn{2}{|c|}{-} & \multicolumn{3}{|c|}{derived parameter, one for PTA}\\
\hline\hline
\multicolumn{9}{|c|}{\textbf{Supermassive Black Hole Binaries}}\\
\hline\hline
$A_{\text{GWB}}$& \multicolumn{3}{|c|}{Strain amplitude} & \multicolumn{2}{|c|}{log-Uniform $[-18, -13]$} & \multicolumn{3}{|c|}{one parameter for PTA} \\

\hline
\end{tabular}}
  \caption{List of noise and GWB parameters used in our analyses, together with their prior ranges.}
  \label{table:priors}
\end{table}

Here we provide further details of our numerical analysis. For the noise analyses, we followed closely the strategies outlined by the NG and IPTA collaborations in their searches for a common spectrum signal in \cite{NANOGrav:2020bcs} and~\cite{Antoniadis:2022pcn}, respectively. We use the datasets released in~\cite{NGnoisedict} for NG12 and in \cite{IPTADR2data} for IPTADR2 (Version B, we use par files with TDB units).

In particular, for both datasets we consider three types of white noise parameters per backend/receiver (per pulsar): EFAC ($E_k$), EQUAD ($Q_k[s]$) and ECORR ($J_k[s]$). The latter is only included for pulsars in the NG12 dataset and for NG 9 years pulsars in the IPTADR2 dataset. Additionally, we included two power-law red noise parameters per pulsar in both datasets: the amplitude at the reference frequency of $\text{yr}^{-1}$, $A_{\text{red}}$, and the spectral index $\gamma_{\text{red}}$. For the IPTA DR2 dataset, we additionally included power-law dispersion measures (DM) errors (see e.g.~\cite{Antoniadis:2022pcn}) (in the single pulsar analysis of PSR J1713+0747 we also included a DM exponential dip parameter following~\cite{Antoniadis:2022pcn}).

In our searches for a GWB, we fixed white noise parameters according to their maximum likelihood a posteriori values from single pulsar analyses (without GWB parameters). In practice, for the NG12 dataset (45 pulsars with more than 3 years of observation time), we used the publicly released white noise dictionary~\cite{NGnoisedict}. For IPTADR2, on the other hand, we built our own dictionary by performing single pulsar analyses for each pulsar with more than 3 years of observation time (we only included those in our search for a GWB, as in~\cite{Antoniadis:2022pcn}, for a total of 53 pulsars). We used the Jet Propulsion Laboratory solar-system
ephemeris DE438, as well as the TT reference timescale BIPM18,
published by the International Bureau of Weights and Measures. 

\begin{figure}[h]
\centering
\includegraphics[width=0.6\columnwidth]{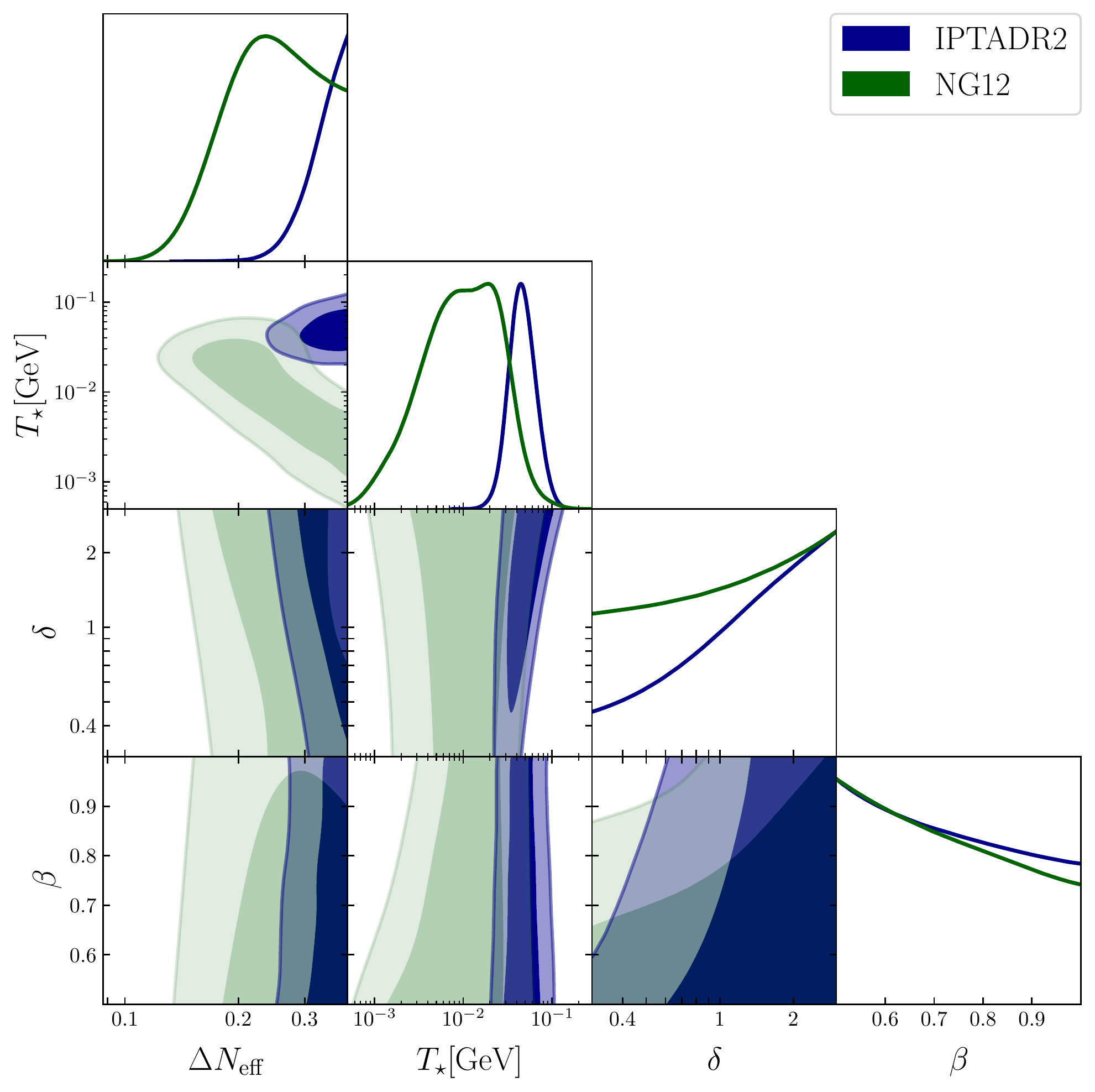}
\includegraphics[width=0.6\columnwidth]{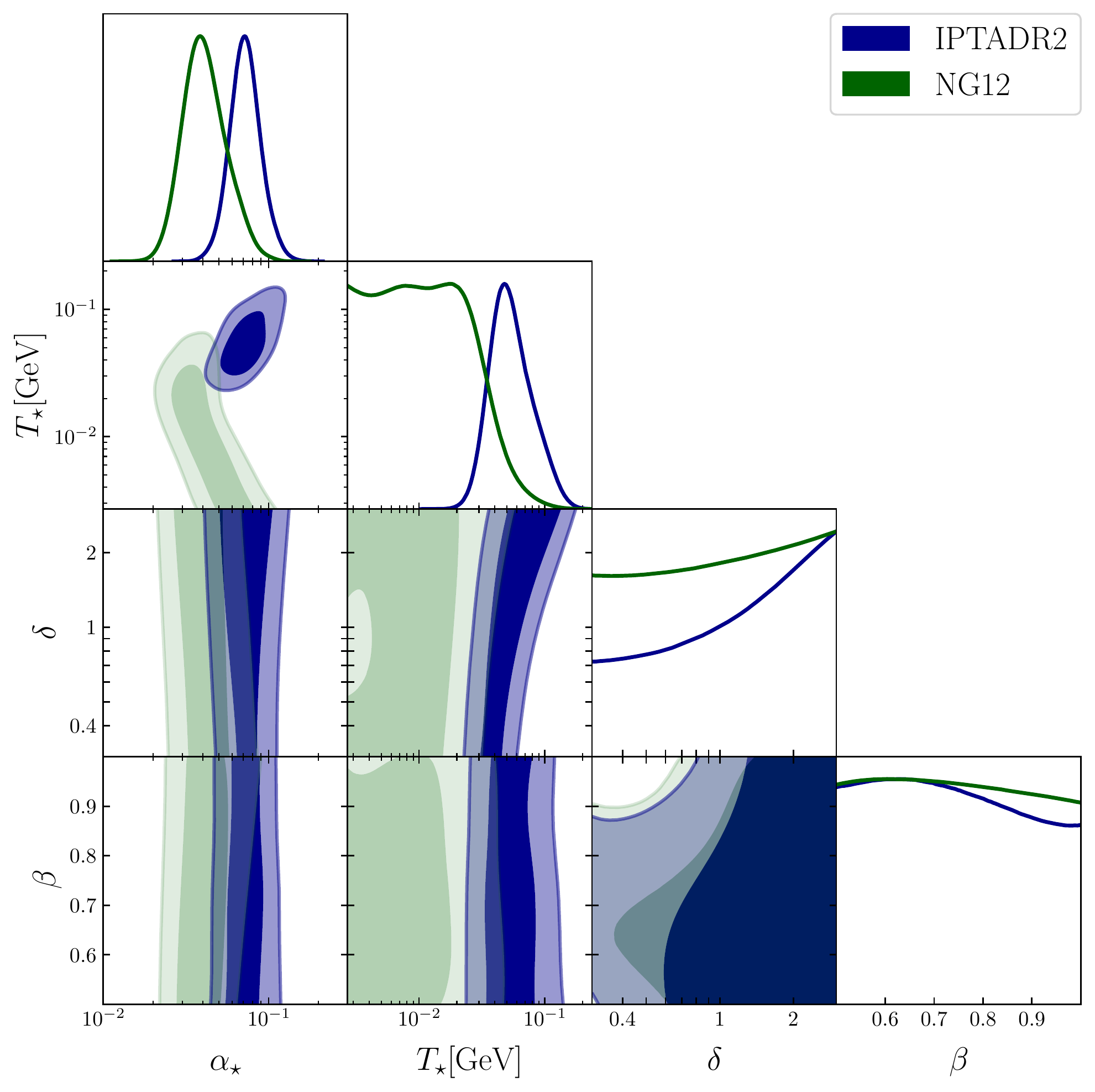}
\caption{One and two-dimensional posterior distributions, with $1\sigma$ and $2\sigma$ contours, of the parameters describing GWs from annihilating DWs. {\it Top:} DW constituents decay to dark radiation. {\it Bottom:} DW constituents decay to SM particles.}
\label{fig:triangleDRSM}
\end{figure}

The choice of priors for the noise parameters in our analyses are reported in Table~\ref{table:priors}, together with the priors for parameters of the GWB from DW annihilation and from SMBHBs. With this strategy and priors for noise parameters, we are able to reproduce the results of \cite{NANOGrav:2020bcs} and~\cite{Antoniadis:2022pcn} for a common-spectrum red-noise process with excellent agreement. We obtain more than $10^6$ samples per each analysis presented in this work and discard $25\%$ of each chain as burn-in.

Following the strategy of~\cite{NANOGrav:2020bcs, Antoniadis:2022pcn}, we use only auto-correlation terms in the Overlap Reduction Function (ORF) in our search, rather than the full Hellings-Downs ORF, to reduce the computational time. We checked in specific cases that this has a minor impact on posterior distributions with the NG12 and IPTADR2 datasets, in agreement with the findings of~\cite{NANOGrav:2020bcs, Antoniadis:2022pcn}.

As described in the main text, we consider two separate cases in our search for GWs from DWs. If DW constituents decay to dark radiation (DR), we express the GW signal in terms of logarithmically sampled parameters $\Delta N_{\text{eff}}$ and $T_\star$, with relevant priors set according to BBN constraints~\cite{Fields:2019pfx} and electron-positron annihilation respectively (the lower and upper bounds on $\Delta N_{\text{eff}}$ and $T_\star$, respectively, are not important). 

If constituents decay to SM particles, we express the GW signal in terms of logarithmically sampled parameters $\alpha_\star$ and $T_\star$, with relevant priors set according to deviation from the radiation dominated background and BBN respectively (the lower and upper bounds on $\alpha_\star$ and $T_\star$, respectively, are again not important). In this case, we also perform a separate analysis expressing the signal in terms of the wall tension $\sigma$ and $T_\star$. The upper boundary of the prior on $\sigma$ is imposed such that there are no deviations from the radiation dominated background. We then obtain posteriors on the derived parameter $\Delta V^{1/4}$ using~\eqref{eq:Tann} and $c=2.2$ (and also $c=4.5$ for string-wall networks with $N=6$~\cite{Kawasaki:2014sqa}) and fixing $g_{*}(T_\star)=15$. A different choice for $g_{*}(T_\star)$ in the range given by lattice calculations does not significantly affects the results, given the very mild dependence of $\Delta V^{1/4}$ on $g_{*}(T_\star)$.

\begin{figure}[]
\centering
\includegraphics[width=0.45\columnwidth]{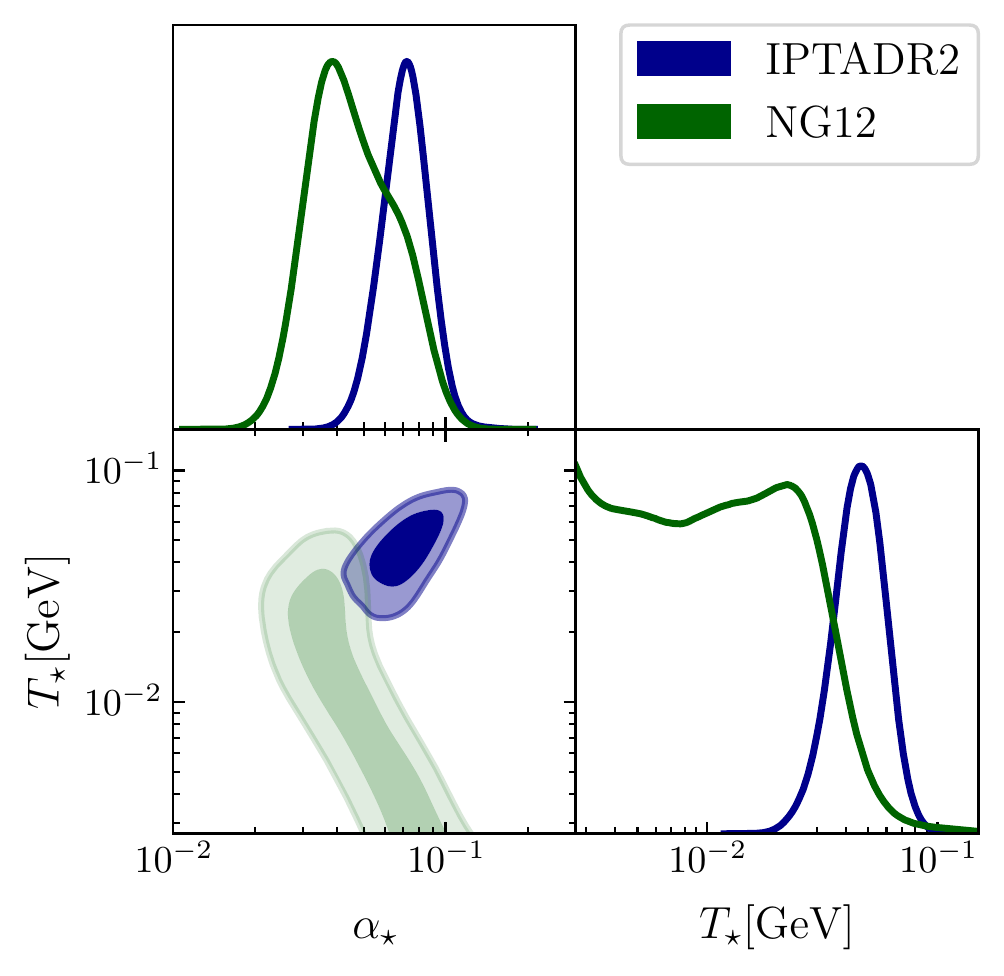}
\caption{One and two-dimensional posterior distributions of the parameters describing GWs from annihilating DWs, fixing $\delta,\beta=1$.}
\label{fig:triangleBD1}
\end{figure}

\begin{figure}[]
\centering
\includegraphics[width=0.6\columnwidth]{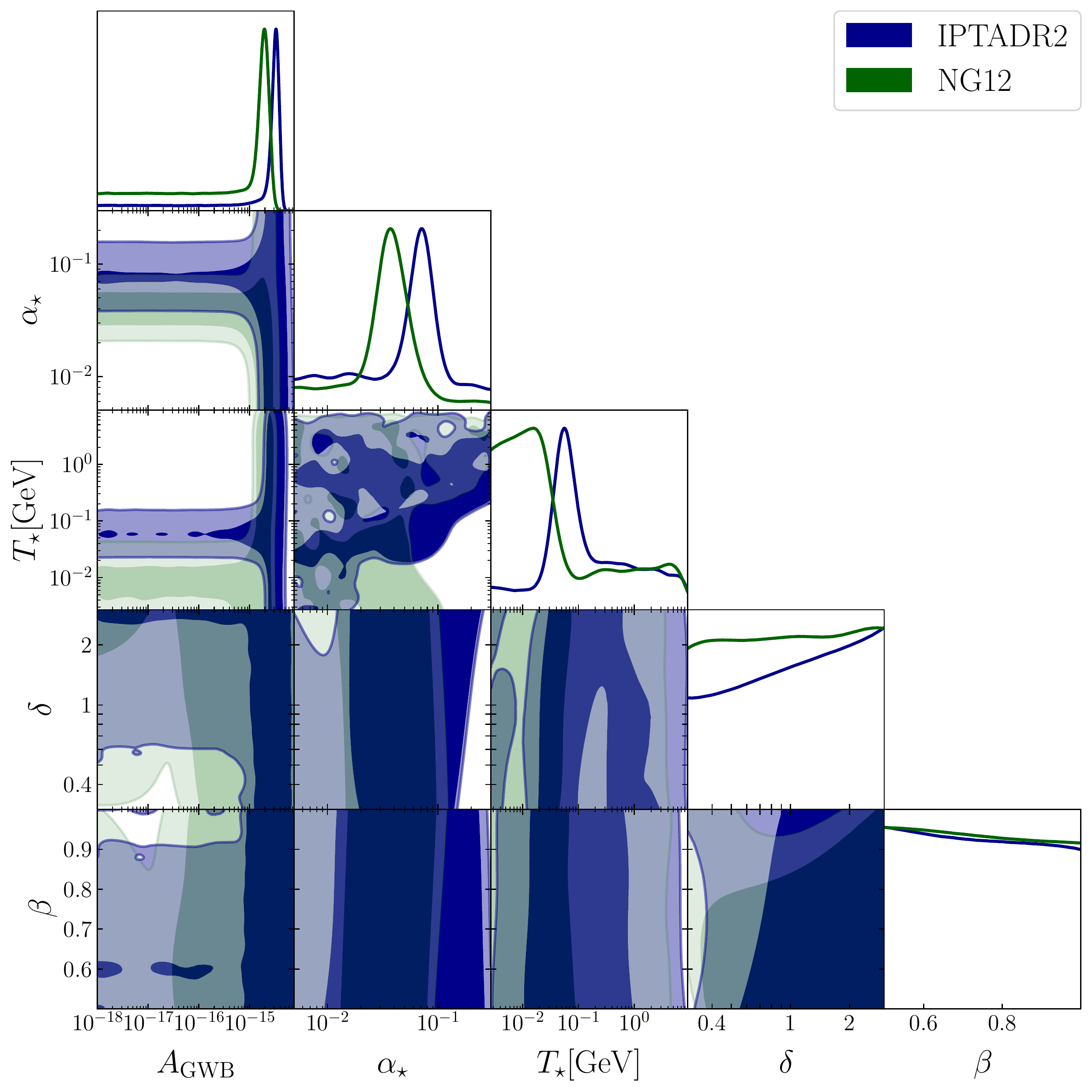}
\includegraphics[width=0.6\columnwidth]{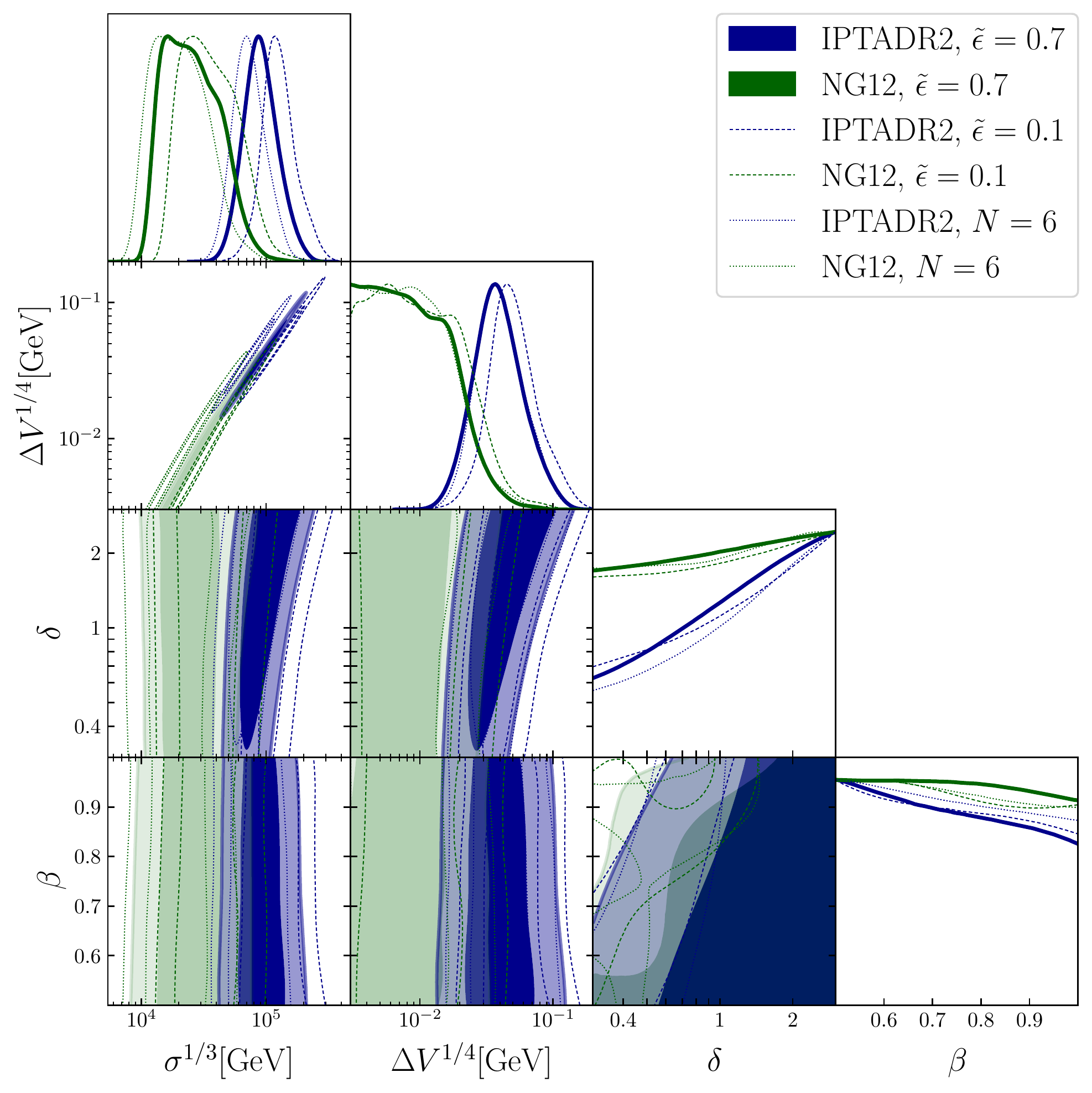}
\caption{One and Two-dimensional posterior distributions, with $1\sigma$ and $2\sigma$ contours, of the parameters describing GWs from annihilating DWs. {\it Top:}
posteriors in the presence of the GWB from SMBHBs, with amplitude $A_\text{GWB}$.{\it Bottom:} posteriors for the case of symmetry breaking by an energy difference $\Delta V$ between vacua. Posteriors on $\Delta V^{1/4}$ have been obtained by means of~\eqref{eq:Tann}, using $c=2.2$ as example value for solid and dashed lines and $c\simeq 4.5$ for dotted lines (from simulations of string-wall networks~\cite{Kawasaki:2014sqa} with $N=3$ and $N=6$ respectively). In both analyses, decay to SM particles has been assumed in imposing priors, see Tables~\ref{table:priors} and~\ref{table:priorsax}.}
\label{fig:triangleBHAX}
\end{figure}

In all cases, we vary the spectral shape parameters $\beta$ and $\delta$, with priors as in Table~\ref{table:priors}. For the SM scenario, we also obtain results with the standard choices $\delta=\beta=1$, to check the minor effect of these parameters on the quality of the fit. Priors for the heavy axion analysis are described in the Appendix below.

Further 1- and 2-dimensional posterior distributions for the DW annihilation and SMBHBs parameters are reported in Figures \ref{fig:triangleDRSM} and \ref{fig:triangleBHAX}. In particular, we observe broad posteriors for the spectral shape parameters $\delta$ and $\beta$, with IPTADR2 very mildly preferring a broader spectral peak. The reference values $\delta,\beta=1$ are both in the $1\sigma$ region of the posteriors in all cases. The effect of fixing these parameters to their unit value is also shown in Fig.~\ref{fig:triangleBD1}, for the SM scenario.

Mean $\pm 2\sigma$ errors, or upper/lower $95\%$ C.L. bounds,  are reported in Table~\ref{table:posteriors} for selected DW parameters.

\begin{table}
\centering
\begin{tabular} {| l | c| c| c}
\hline\hline
 \multicolumn{1}{|c|}{ \textbf{Parameter}} & \multicolumn{1}{|c|}{~~~\textbf{IPTADR2}~~~} & \multicolumn{1}{|c|}{~~~~\textbf{NG12}~~~~} \\
 \hline
 \multicolumn{3}{|c|}{\textbf{DR scenario}}\\
 \hline
$\log_{10} T_\star/\text{GeV}$ & $-1.34^{+0.31}_{-0.29}$ & $\leq -1.29$ \\
$\log_{10} \Delta N_{\text{eff}}$ & $\geq -0.58$ & $\geq -0.81$ \\
\hline
 \multicolumn{3}{|c|}{\textbf{SM scenario}}\\
 \hline
$\log_{10} T_\star/\text{GeV}$ & $-1.27^{+0.35}_{-0.30}$ & $\leq -1.39$ \\
$\log_{10}\alpha_\star$ & $-1.14^{+0.20}_{-0.18}$ & $-1.39^{+0.28}_{-0.24}$ \\
\hline
$\log_{10}\sigma^{\frac{1}{3}}/\text{GeV}$ & $4.96^{+0.29}_{-0.26}$  & $4.41^{+0.40}_{-0.36}$\\
$\log_{10}\Delta V^{\frac{1}{4}}/\text{GeV}$ & $-1.43^{+0.40}_{-0.35}$ & $\leq -1.59$ \\
\hline
\end{tabular}
  \caption{The mean $\pm 2\sigma$ error (lower/upper bounds) of DW parameters in the scenarios considered in this work.}
  \label{table:posteriors}
\end{table}

\section{The Heavy Axion \label{sec:axion}}

Here we provide more details on the heavy axion origin of DWs. We consider a global $U(1)$ symmetry, spontaneously broken in the early Universe after reheating. To fix ideas, this can arise from a complex scalar field $\Phi$ with potential $V(\Phi)=\lambda(\lvert\Phi\rvert^2-v^2/2)^2$. The axion field $a$ is the resulting Goldstone boson with a periodicity $2\pi v$. At this stage, topological defects, known as cosmic strings and mostly made of the massive radial mode, appear. In the presence of a coupling to a confining dark gauge sector $\mathcal{H}$, e.g. a $SU(n)$ gauge theory with no massless fermions, $a$ receives a periodic potential of the form:
\begin{equation}
    V_{\mathcal{H}}(a)=m_a^2F_a^2\left[1-\cos\left(\frac{a}{F_a}\right)\right],
\end{equation}
where $m_a\simeq \kappa_\mathcal{H}\Lambda^2_\mathcal{H}/F_a$ and $F_a\equiv v/N$ is the axion decay constant.  $N$ is an integer that depends on the matter content of the dark sector that is charged under the $U(1)$ symmetry. We included here the factor $\kappa_\mathcal{H}\lesssim 1$ that can arise e.g. in case that this sector includes a light fermion of mass $m_q$, giving $\kappa_\mathcal{H}\sim \sqrt{m_q/\Lambda_{\mathcal{H}}}$, see  \cite{ZambujalFerreira:2021cte} (in the main text we set $\kappa_\mathcal{H}\simeq 1$  to estimate $\Lambda_\mathcal{H}$ from Fig.~\ref{fig:axion}). We focus on the case $N>1$ (which arises e.g. when there is more than one vector-like fermion pair charged under the $U(1)$), that leads to a residual $\mathbb{Z}_N$ symmetry. The latter is spontaneously broken at temperatures around $\mathcal{H}$ confinement, and a long-lived network of DWs attached to the previously formed strings, with $N$ walls attached to each string~(see~\cite{Vilenkin:2000jqa} for a review), is formed. The network rapidly enters the scaling regime in the absence of thermal friction.

\begin{table}
\resizebox{\textwidth}{!}{%
\begin{tabular} {| l | c| c| c| c|c|c|c|c|}
\hline\hline
 \multicolumn{1}{|c|}{ Parameter} & \multicolumn{3}{|c|}{~~~~~Description~~~~~} & \multicolumn{2}{|c|}{~~~~Prior~~~~} & \multicolumn{3}{|c|}{~~Comments~~} \\
\hline\hline
\multicolumn{9}{|c|}{\textbf{Domain Wall Annihilation (Heavy Axion)}}\\
\hline\hline
$T_\star [\text{GeV}]$& \multicolumn{3}{|c|}{Annihilation temperature} & \multicolumn{2}{|c|}{log-Uniform $[\log_{10}(0.0027), 10]$} & \multicolumn{3}{|c|}{one parameter for PTA} \\
$F_{a} [\text{GeV}]$& \multicolumn{3}{|c|}{Axion decay constant} & \multicolumn{2}{|c|}{log-Uniform $[2, 9]$} & \multicolumn{3}{|c|}{one parameter for PTA} \\
$m_a [\text{GeV}] $& \multicolumn{3}{|c|}{Axion mass} & \multicolumn{2}{|c|}{log-Uniform $[\log_{10}(4\cdot 10^{-4}), 4]$} & \multicolumn{3}{|c|}{one parameter for PTA} \\
$\beta $& \multicolumn{3}{|c|}{High frequency spectral index} & \multicolumn{2}{|c|}{Uniform $[0.5, 1]$} & \multicolumn{3}{|c|}{one parameter for PTA} \\
$\delta $& \multicolumn{3}{|c|}{Spectrum width} & \multicolumn{2}{|c|}{log-Uniform $[0.3, 3]$} & \multicolumn{3}{|c|}{one parameter for PTA} \\
\hline
$\mu_b$ [\text{GeV}]& \multicolumn{3}{|c|}{Size of misaligned potential} & \multicolumn{2}{|c|}{-} & \multicolumn{3}{|c|}{derived parameter, one for PTA} \\
\hline
\end{tabular}}
  \caption{GWB parameters used in our analysis of GWs from heavy axion DW annihilation, together with their prior ranges.}
  \label{table:priorsax}
\end{table}

\begin{figure}[]
\centering
\includegraphics[width=0.45\columnwidth]{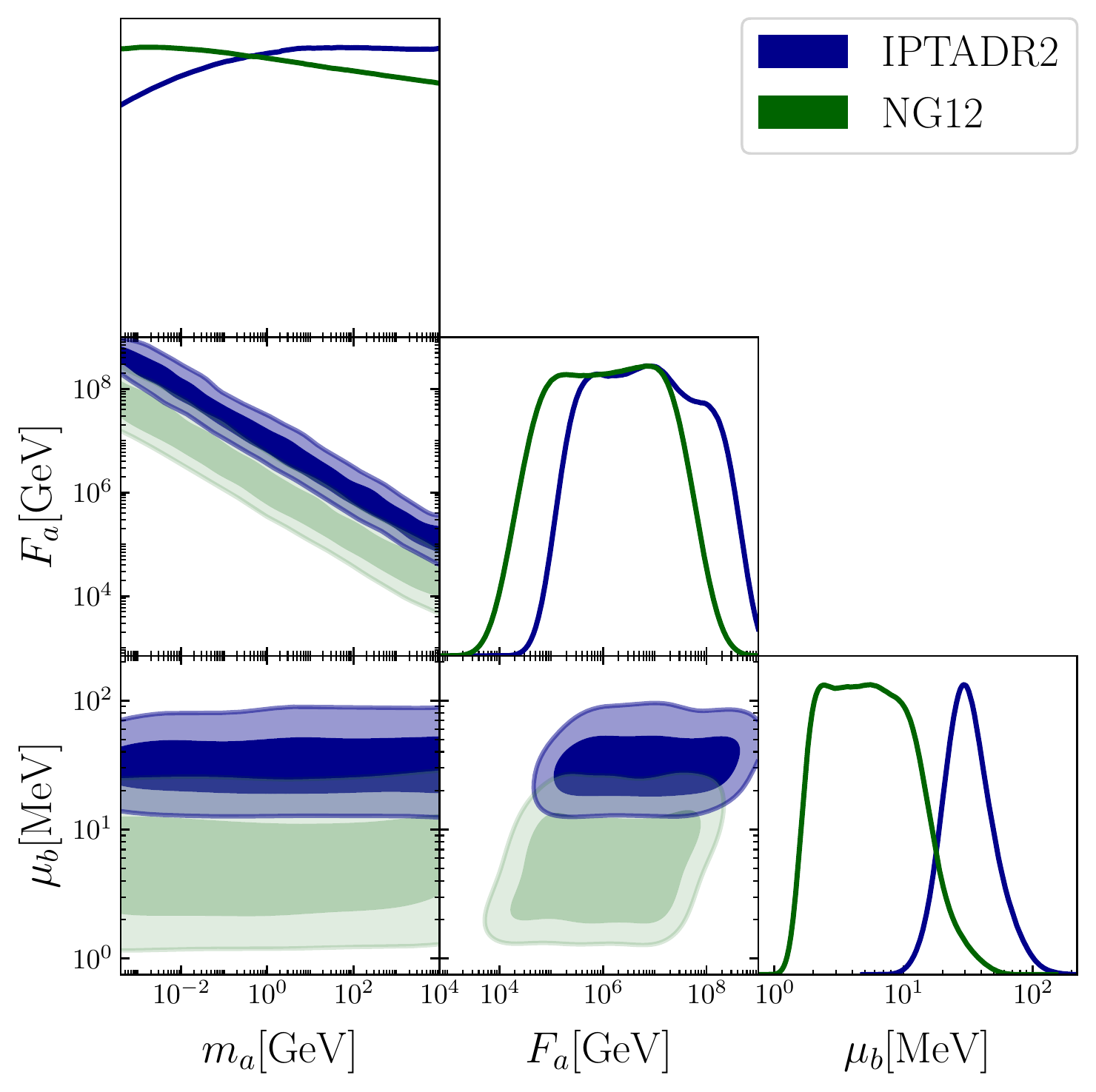}
\caption{One and Two-dimensional posterior distributions, with $1\sigma$ and $2\sigma$ contours, of the parameters describing GWs from heavy axion DWs. The posteriors on the size of the gap energy $\mu_b^{1/4}$ are obtained using~\eqref{eq:Tann} and~\eqref{eq:DeltaV}, for the case $N=6$, i.e. fixing $c=4.48$ according to~\cite{Kawasaki:2014sqa}.}
\label{fig:triangleAX}
\end{figure}

We consider two possibilities to induce DW annihilation (see~\cite{ZambujalFerreira:2021cte} for more details). First, the existence of another confining sector at a scale $\Lambda_c\ll \Lambda_\mathcal{H}$. Second, the presence of high scale $U(1)$-breaking effects which manifest themselves either via higher-dimensional operators in $\Phi$ or via direct non-perturbative contributions to the axion potential (see e.g.~\cite{Dine:1986bg, Kamionkowski:1992mf}). In both cases, the axion potential is corrected by a term of the form
\begin{eqnarray}
	V_b = -\mu_b^4 \cos \left(\frac{M}{N} \frac{a}{F_a} -\delta \right) \,,
\end{eqnarray}
where $M$ is an integer. In the first case, $\mu_b\simeq \sqrt{\kappa_c}\Lambda_c$ where $\kappa_c\simeq \sqrt{m_c/\Lambda_c}$ is again the mass of the lightest state below $\Lambda_c$ in the additional confining sector. In the second case, $\mu_b\simeq c_n^{1/4}(N F_a/\Lambda)^{n/4}\Lambda$ for operators of dimension $n$ with coefficients $c_n$ and suppressed by a high scale $\Lambda$, or $\mu_b\simeq e^{-S/4}\Lambda$ for non-perturbative contributions (see e.g. discussion in~\cite{ZambujalFerreira:2021cte}). The phase $\delta$ is a generic misalignment with respect to $\mathcal{H}$. When $M=1$ or is co-prime with $N$, $V_b$ lifts the degeneracy of the $N$ minima. When $\mu_b^4\ll m_a^2 F_a^2$, the energy difference between two neighboring minima is estimated as 
\begin{equation}
\label{eq:DeltaV}
    \Delta V\simeq \mu_b^4\left[1-\cos\left(\frac{2 M\pi}{N}\right)\right],
\end{equation}
and the temperature $T_\star$ can be determined by means of~\eqref{eq:Tann}. 

In our numerical search, we express the GW relic abundance in \eqref{eq:omegadw} and frequency in \eqref{eq:pfreq} in terms of three parameters ($T_\star, F_a, m_a$) in order to perform a comparison with other searches and experiments. We report priors for those parameters in Table~\ref{table:priorsax}. The lower boundaries of the prior ranges for $T_\star$ and $m_a$ are chosen according to BBN constraints~\cite{Depta:2020zbh}. We then obtain the size of the misaligned potential $\mu_b$ as a derived parameter, by means of~\eqref{eq:Tann} and~\eqref{eq:DeltaV}. We fix $M=1, N=6$ as example values and correspondingly set $c=4.48$ according to the simulations \cite{Kawasaki:2014sqa}. We also vary $\beta, \delta$ as in Table~\ref{table:priors}. Posterior distributions are shown in Fig.~\ref{fig:triangleAX}.

Let us now discuss the possibility that $\Delta V$ originates from QCD (this is not required by PTA data). This corresponds to setting $\mu_b\simeq 75.5~\text{MeV}$, for which one finds $\Delta V\simeq 80~\text{MeV}$, for example values $N=3, M=1$, and $\Delta V\simeq 60~\text{MeV}$, for $N=6, M=1$. These values fit nicely inside the marginalized $2\sigma$ posteriors for $\Delta V$ inferred from IPTADR2 (and may also fit NG data well if future noise analyses of NG12 data find better agreement with IPTADR2).

Of course, if there is no other axion field, $a$ needs to solve the strong CP problem and thus $\mathcal{H}$ and QCD need to be aligned down to $\delta\lesssim 10^{-10}$. This can be realized in so-called heavy QCD axion scenarios (see e.g.~\cite{Dimopoulos:2016lvn,  Agrawal:2017ksf, Hook:2019qoh, Gherghetta:2020keg} for recent work). However, such alignment is typically ensured by means of a symmetry (e.g. $\mathbb{Z}_2$), and thus it is often the case that $M=N$ and QCD cannot actually induce DW annihilation. If this is the case, annihilation needs to occur due to other sources of $U(1)$ breaking, such as those considered above.

On the other hand, if a second axion $b$ which couples only (mostly) to QCD and solves the strong CP problem exists, the two sectors can be generically misaligned and unrelated (see~\cite{Draper:2018tmh} for a discussion). This case then appears more promising to realize DW annihilation from QCD instantons. Let us mention that GWs from multi-axion DW networks have been considered in the so-called clockwork model~\cite{Higaki:2016jjh} (see also~\cite{Long:2018nsl}). Beyond a specific form of the axion potentials, these works also assumed that the $U(1)$ symmetries generating the axion fields are broken at the same scale, and concluded that the network is long-lived only when the number of axions is at least three. It would be interesting to extend the analysis of~\cite{Higaki:2016jjh} to the more general two-axion string-wall network considered in our work and to understand whether an additional axion is required in our case as well.

Whenever the axion $a$ couples to QCD, it can efficiently decay to SM gluons or photons, as described in~\cite{Aloni:2018vki}, with a decay rate $\Gamma_{a\rightarrow gg, \gamma\gamma}\propto m_a^3/F_a^2$. We verified that for most of the parameter space in Fig. \ref{fig:axion}, apart from a small corner in the upper left part, such decays are efficient at $T_\star$.

\clearpage

\bibliography{biblio}
\bibliographystyle{JHEP}

\end{document}